# Multiferroicity and skyrmions carrying electric polarization in GaV$_4$S$_8$

E. Ruff,[1]* S. Widmann,[1] P. Lunkenheimer,[1] V. Tsurkan,[1,2] S. Bordács,[3] I. Kézsmárki[1,3], A. Loidl[1]

[1]Experimental Physics V, Center for Electronic Correlations and Magnetism, University of Augsburg, 86135 Augsburg, Germany
[2]Institute of Applied Physics, Academy of Sciences of Moldova, Chisinau MD-2028, Republic of Moldova
[3]Department of Physics, Budapest University of Technology and Economics and MTA-BME Lendület Magneto-optical Spectroscopy Research Group, 1111 Budapest, Hungary

*Corresponding author. E-mail: eugen.ruff@physik.uni-augsburg.de

**Skyrmions are whirl like topological spin objects with high potential for future magnetic data storage. It is a fundamental question, relevant for both basic research and application, if a ferroelectric (FE) polarization can be associated with their magnetic texture and if these objects can be manipulated by electric fields. Here, we study the interplay between magnetism and electric polarization in the lacunar spinel GaV$_4$S$_8$, which undergoes a structural transition associated with orbital ordering at 44 K and reveals a complex magnetic phase diagram below 13 K, including a ferromagnetic (FM), cycloidal, and Néel-type skyrmion lattice (SkL) phase. We found that the orbitally ordered phase of GaV$_4$S$_8$ is FE with a sizable polarization of ~1 µC/cm$^2$. Moreover, we observed spin-driven excess polarizations in all magnetic phases and, hence, GaV$_4$S$_8$ hosts three different multiferroic phases with coexisting polar and magnetic order. These include the SkL phase where we predict a strong spatial modulation of the FE polarization close to the skyrmion cores. By taking into account the crystal symmetry and spin patterns of the magnetically ordered phases, we identify the exchange striction as the main microscopic mechanism behind the spin-driven FE polarization in each multiferroic phase. Since GaV$_4$S$_8$ is unique among the known SkL host materials due to its polar crystal structure and the observed strong magnetoelectric effect, this study is an important step towards the non-dissipative electric-field control of skyrmions.**

## INTRODUCTION

In recent years, magnetic skyrmions, whirl-like spin textures with non-trivial topology and typical sizes of 1-100 nm, have attracted tremendous interest from both an academic and technological point of view (*1,2,3,4*). Besides the formation of individual skyrmions, these objects can be arranged in skyrmion lattices (SkL), representing a new form of magnetic order (*1,2*). Due to their topology and nanometric size, skyrmions are stable against external perturbations (*5*). Since each skyrmion is associated with a quantized magnetic flux, their creation and annihilation must be related to singular magnetic defects, which can be regarded as an analogue of magnetic monopoles (*6,7*). Metallic chiral magnets, where skyrmions can be moved over macroscopic distances by low electrical currents, can be exploited for spintronic applications such as race-track memories (*3*). In addition, insulating skyrmion host materials have been discovered recently (*8,9*) raising the possibility of multiferroic behavior in these systems. Multiferroicity, the coexistence of magnetic and polar orders, may favor the manipulation of the magnetic pattern via electrical fields without Joule heating (*10,11*). Such a non-dissipative electronic control of nanometric magnetic objects is very promising for applications in new generations of magnetic memories and spintronic devices. Indeed, shortly after the first observation of skyrmions in multiferroic $Cu_2OSeO_3$ (*8*), the electric-field control of the SkL has also been demonstrated in this compound (*12,13*). The dielectric response and emergent electrodynamics of skyrmions have also been extensively studied experimentally and theoretically both in itinerant and insulating magnets (*5,7,14,15,16,17,18*)

Lacunar spinels, ternary chalcogenides of composition $AM_4X_8$ ($A$ = Ga and Ge; $M$ = V, Mo, Nb, and Ta; $X$ = S and Se), represent an interesting class of transition-metal compounds with weakly linked molecular units, cubane $(M_4X_4)^{n+}$ and tetrahedral $(AX_4)^{n-}$, as structural building blocks (*19,20,21,22*). Recently, a plethora of correlation effects has been reported for $AM_4X_8$ lacunar spinels, including pressure-induced superconductivity (*23*), bandwidth-controlled metal-to-insulator transition (*24,25*), large negative magnetoresistance (*26*), a two-dimensional topological insulating state (*27*), resistive switching via an electric-field induced transition (*28,29,30*), emergence of orbitally driven ferroelectricity (*31*), and, most interestingly, an extended Néel-type SkL phase (*9*).

The vanadium-derived lacunar spinel, $GaV_4S_8$ is a magnetic semiconductor which has a non-centrosymmetric cubic structure (point symmetry group $T_d$) at room temperature (*32,33*). It consists of a network of weakly coupled $(V_4S_4)^{5+}$ cubane units forming a face centered cubic (fcc) lattice, separated by $(GaS_4)^{5-}$ tetrahedra. Each vanadium $V_4$ cluster carries a single local spin $S$ = ½ (*33*). $GaV_4S_8$ undergoes a cubic-to-rhombohedral structural phase transition at $T_{JT}$ = 44 K and magnetic ordering at $T_C$ = 13 K (*22,34,35*). The structural phase transition has been identified as a Jahn-Teller derived orbital ordering characterized by an elongation of the $V_4$ tetrahedra along one of the four crystallographic $\langle 111 \rangle$ directions (*33,36*) (see inset of Fig. 1B) leading to a low-temperature rhombohedral $C_{3v}$ point group symmetry (*32,33*). Ref. 9 has revealed a complex magnetic phase diagram of $GaV_4S_8$, with a cycloidal and SkL phase embedded within the FM state. Skyrmion lattices have recently been observed in various magnets with chiral structure (*1,2,8,37*). Most interestingly, in $GaV_4S_8$ novel Néel-type skyrmions, which are composed of spin cycloids and carry a monopole moment (*38*), have been identified (*9*).

Here we report a detailed study of the FE properties of $GaV_4S_8$ as function of temperature and magnetic field, supplemented by specific heat and magnetic susceptibility measurements. We provide experimental evidence for the onset of ferroelectric order with sizable polarization at the Jahn-Teller transition. In addition, we find that the FE polarization is enhanced when crossing the magnetic phase boundaries, leading to different polarizations in the collinear FM, cycloidal, and SkL phase. Based on the spin patterns in the three magnetically ordered phases and the symmetry of the $V_4$-$V_4$ bonds in the rhombohedral phase, we identify exchange striction as the main microscopic origin of all spin-driven FE polarizations in $GaV_4S_8$ and estimate the corresponding exchange-striction parameters from the variation of the polarization between the different magnetic phases.

## RESULTS

**Sample characterization.**

The GaV$_4$S$_8$ single crystal studied in the present Report has been characterized by means of magnetic susceptibility and specific heat measurements (for details see the Supplementary Materials). Both the structural and the magnetic phase transitions are clearly manifested in the anomalies of these thermodynamic quantities at $T_{JT}$ = 44.0 K and $T_C$ = 12.7 K, respectively. The structural phase transition as seen in the susceptibility and dielectric measurements is accompanied by a hysteresis loop on heating and cooling, similar to the findings in GeV$_4$S$_8$ (*39*) (see the Supplementary Materials).

Figure 1A shows the temperature dependence of the real part of the dielectric constant measured at selected frequencies. A significant anomaly appears at the Jahn-Teller transition, which substantially depends on frequency. The strong suppression of the maximum dielectric constant at the structural phase transition resembles the behavior commonly observed in order-disorder ferroelectrics (*40*). At high temperatures and low-frequencies, the dielectric constant is strongly influenced by extrinsic Maxwell-Wagner-like contributions, which arise from contact and surface effects and yield values $> 10^3$ at room temperature (*41,42*). Such extrinsic contributions to the low-frequency dielectric constant are still present at $T_{JT}$; hence, we restrict ourselves to relatively high frequencies where intrinsic dielectric constants are measured and Maxwell-Wagner contributions are shifted to high temperatures above 100 K (see the Supplementary Materials). When approaching the Jahn-Teller transition from the cubic phase, the increase of the dielectric constant, as measured at 2.5 GHz and documented in the inset of Fig. 1A, suggests the presence of strong polar fluctuations. It seems quite natural to correlate these polar fluctuations with orbital fluctuations via a dynamic Jahn-Teller effect. In this case the primary order parameter is the Jahn-Teller distortion which gives rise to the FE polarization due to the non-centrosymmetric crystal structures of the low and high-temperature phases. The onset of magnetic order is hardly visible in the temperature dependence of the dielectric constant. It is worth mentioning that the temperature dependence of the dielectric constant in the FE and isostructural GeV$_4$S$_8$ also shows very unusual behavior (*31*).

In GaV$_4$S$_8$ the FE polarization along the $\langle 111 \rangle$ direction, determined from pyrocurrent measurements, abruptly appears at the structural transition (see Fig. 1B), in agreement with the sharp anomaly observed in the heat capacity (see the Supplementary Materials). The polarization saturates somewhat below 0.6 µC/cm$^2$, which is significantly larger than the polarization in spin-driven multiferroics (*43,44,45*), but still a factor of 50 lower than in canonical perovskite ferroelectrics (*46*). This value is close to the FE polarization found in the related lacunar spinel GeV$_4$S$_8$ (*31*). In Fig. 1B we show the elongation of the vanadium tetrahedra along $\langle 111 \rangle$ driven by concomitant orbital order originally proposed in Refs. (*33,36*), which is responsible for the macroscopic polarization. According to this scenario GaV$_4$S$_8$ is an orbital-order driven FE, where the lifting of orbital degeneracy at the Jahn-Teller transition induces FE polarization.

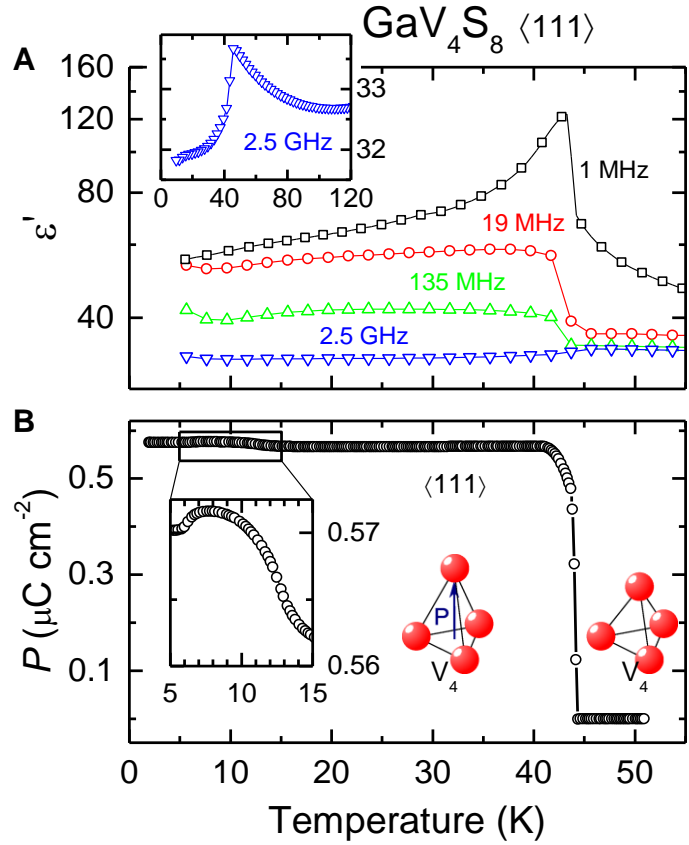

**Fig. 1. Characterization of dielectric properties.** (**A**) Real part of the dielectric constant ε' measured at frequencies between 1 MHz (open squares) and 2.5 GHz (open triangles down). The electric field E was applied parallel to the crystallographic ⟨111⟩ direction. The inset shows an enlarged view of the measurements at 2.5 GHz. (**B**) Electric polarization **P** vs. temperature measured parallel to the ⟨111⟩ direction. The elongation of the V$_4$ units along ⟨111⟩ is schematically indicated and contrasted with the orbitally degenerate high-temperature phase. The inset shows the excess polarization at the onset of magnetic order on an enlarged scale.

A closer inspection of Fig. 1B reveals additional features related to the magnetic phase transitions. The inset in Fig. 1B provides an enlarged view of the low-temperature polarization, which shows an increase upon entering into the cycloidal phase at 12.7 K and a small decrease close to 6 K when the low-temperature FM order is finally established (*9*). This finding documents that in GaV$_4$S$_8$, in addition to the orbital-derived FE polarization of approximately 6000 μC/m$^2$, an excess spin-driven polarization of the order of 100 μC/m$^2$ appears in the magnetic phases. In contrast, in GeV$_4$S$_8$ the magnetically ordered phase shows no excess but rather a reduction of polarization (*31*). In the remaining part of this article, we focus on the most important and fascinating aspect of these results, namely the multiferroic nature of the different magnetically ordered phases, including the SkL phase, where the skyrmions are dressed with local FE polarization. Specifically, it seems important to clarify the origin of the excess polarizations, which will be done in the theory section of this manuscript.

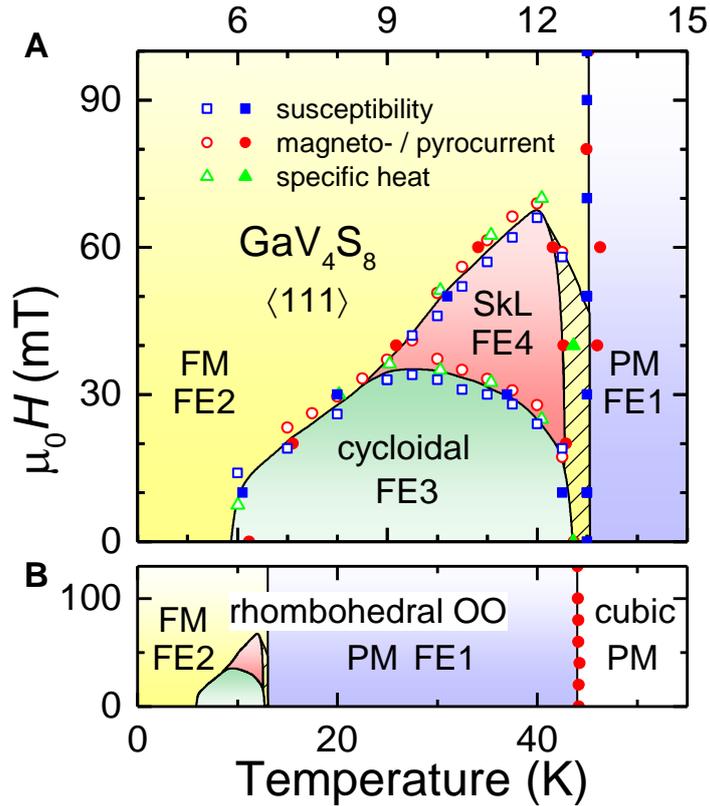

**Fig. 2. Magnetic field versus temperature phase diagram of GaV$_4$S$_8$ for magnetic fields applied along one of the cubic ⟨111⟩ axes.** (**A**) Low-temperature region below $T$ = 15 K. The magnetic phase boundaries correspond to the unique structural domain whose easy axis is parallel to the magnetic field (for details see text). The SkL phase (red area) and the cycloidal phase (green area) are embedded within the FM phase (yellow area). Above $T_C$ = 12.7 K the material is paramagnetic (PM). The phase boundaries as determined from magnetization, polarization (pyro- and magneto-current) and specific heat measurements are indicated by squares, circles, and triangles, respectively. Full symbols correspond to temperature scans, while open symbols represent magnetic-field scans. The hatched area characterizes the regime between the PM and the cycloidal/SkL phase, where the exact spin structure is unknown. (**B**) Complete phase diagram extending beyond the Jahn-Teller transition. Below $T_{JT}$, within the orbitally ordered (OO) phase, four different FE states (FE1 – FE4) are found, each characterized by well distinguishable polarization.

**Phase diagram of GaV$_4$S$_8$**

In Fig. 2 we present the phase diagram of GaV$_4$S$_8$ based on magnetization, magneto-, and pyrocurrent as well as on specific-heat measurements for magnetic fields applied along the crystallographic ⟨111⟩ direction (a phase diagram based on magnetization measurements alone was published in ref. 9). In GaV$_4$S$_8$ due to easy-axis exchange anisotropy, the skyrmion cores are always aligned parallel to the ⟨111⟩ easy axis and they are not oriented by low external magnetic fields. Therefore, when the magnetic field is applied along ⟨111⟩, one magnetically favorable domain coexists with three unfavorable domains, whose cycloidal and SkL states persist up to higher fields (9). The finite electric polarization measured without poling electric fields indicates unequal domain population. Moreover, since its value does not change after several warming-cooling cycles, we assume that a unique domain, i.e. essentially a monodomain polarization, is stabilized in the present experiment. Therefore, in Fig. 2 we focus on the low-field phase diagram specific to this domain, neglecting the cycloidal and SkL phases occurring at higher fields in the magnetically unfavorable domains (*9*).

Figure 2A presents the magnetic-field versus temperature phase diagram of $GaV_4S_8$ for $T < 15$ K, while Fig. 2B shows the complete H-T phase diagram extending beyond the Jahn-Teller transition. Below the magnetic phase transition, occurring at $T_C = 12.7$ K, two magnetic-phase pockets, the cycloidal and the SkL phases, are embedded in the FM phase (*9*). In zero magnetic field, the FM, the cycloidal and the orbitally ordered paramagnetic (PM) phases are all FE as evidenced by Fig. 1B (denoted in Fig. 2 as FE2, FE3, and FE1, respectively). The orbitally disordered phase above $T_{JT} = 44$ K is a PM semiconductor without FE polarization.

To further elucidate the phase diagram (Fig. 2A), a representative set of magnetization, specific heat and polarization curves as function of temperature and magnetic field is shown in Fig. 3. At the phase boundaries clear anomalies can be identified in the field dependent (Figs. 3A,C,E), as well as in the temperature dependent scans (B,D,F) of all three quantities. The only exception is the temperature dependence of the heat capacity (Fig. 3D) which reveals no anomaly when passing from the cycloidal to the FM phase. This may be due to the fact that the entropy change between the cycloidal and the ferromagnetic states is expected to be very small due to the long wavelength of the cycloid. The position of all observed anomalies are included in Fig. 2A, which are in perfect agreement with each other and the published phase boundaries (*9*) and, thus, convincingly document the complexity of the low-temperature phase diagram of $GaV_4S_8$ including a relatively extended SkL phase which reveals spontaneous FE polarization. It has to be noted that the excess spin-driven polarization $\Delta P$ at the magnetic phase boundaries, as presented in Figs 3E and 3D, appears on top of a much larger orbital-order derived polarization (see Fig. 1B).

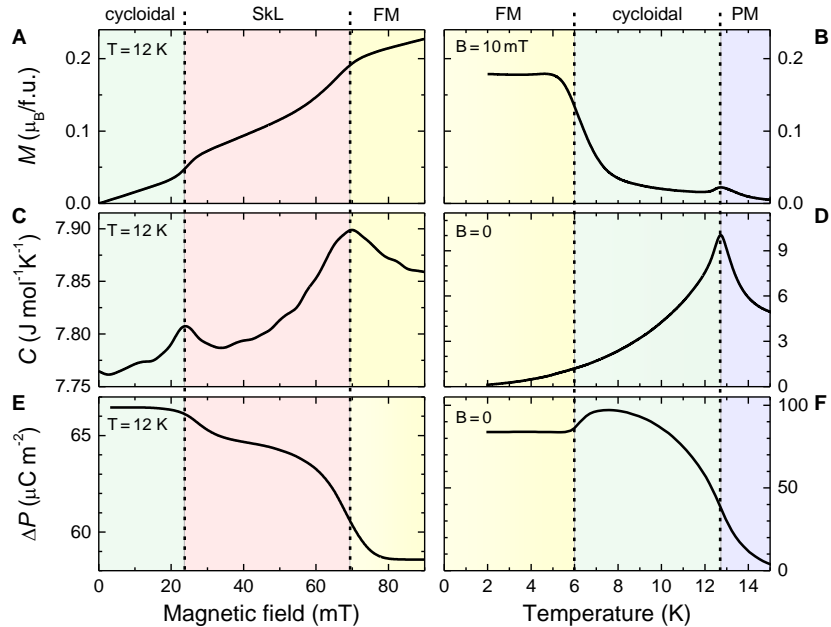

**Fig. 3. Anomalies in magnetization, specific heat, and FE polarization at the magnetic phase boundaries.** (**A**),(**C**),(**E**) Magnetic-field dependence of magnetization (*M*), specific heat (*C*), and isothermal polarization ($\Delta P$) with $H||\langle 111 \rangle$. (**B**),(**D**),(**F**) Temperature dependence of magnetization, specific heat, and zero-field polarization. Background colours represent the different magnetic phases following the colour code used in Fig. 2.

**Ferroelectricity in the magnetic phases**

To obtain more detailed information about the different FE phases, including the SkL phase, we studied the temperature (Fig. 4) and field dependence (Fig. 5) of the FE polarization deduced from pyro- and magnetocurrent measurements, respectively. In all scans presented in Figs. 3-5, only the excess polarizations of the magnetic phases are shown, which are less than 2% of the orbitally-derived ferroelectricity emerging at $T_{JT}$ (see inset in Fig. 1B).

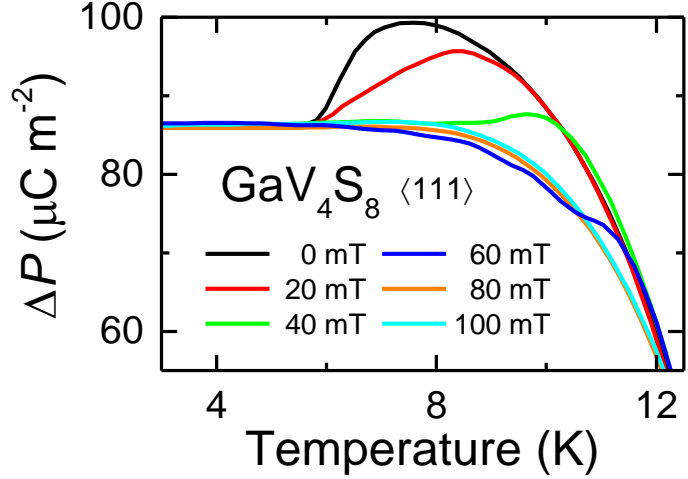

**Fig. 4. Ferroelectric polarization of $GaV_4S_8$ determined from pyrocurrent measurements.** The figure shows the polarization as function of temperature measured in various magnetic fields between 0 and 100 mT. Only the excess polarization $\Delta P$, arising when entering the magnetic phases at $T_C$ = 12.7 K, is shown.

Fig. 4 shows the temperature dependence of the excess polarization $\Delta P$ in various magnetic fields between 0 and 100 mT, while Fig. 5 presents the polarization versus magnetic field at various temperatures between 2 and 13 K. The steps observed in the field and temperature dependence of the polarization, corresponding to peaks in the magneto- and pyrocurrent curves, are indicated in Fig. 2A by open and closed circles, respectively. They nicely match the phase boundaries determined from specific heat and magnetization data (see the Supplementary Materials). All magnetically ordered phases are characterized by different excess polarizations. In fields of $\mu_0 H$ = 100 mT, the material undergoes a PM to FM transition at $T_C$ = 12.7 K without any intermediate phase, and the excess polarization in the FM phase reaches the saturation value, $\Delta P_{FM}$ ~ 86 $\mu C/m^2$, at the lowest temperatures (see Fig. 4). The curves measured in $\mu_0 H$ = 0 mT and 20 mT, following a steep increase at 13 K, exhibit an additional drop at about 6 K, where the transition from the cycloidal to FM phase takes place (see Fig. 2). Therefore, the polarization in the cycloidal phase is enhanced by approximately 14 $\mu C/m^2$ compared to the purely FM collinear spin arrangement, resulting in a total excess polarization of $\Delta P_{cyc}$ ~ 100 $\mu C/m^2$. The SkL to FM phase boundary is also accompanied by small but significant anomalies as revealed by the polarization curves measured in $\mu_0 H$ = 40 and 60 mT (Fig. 4). The scan at 40 mT, only crossing the SkL state between 8 and 13 K, reveals an extra contribution of about 5 $\mu C/m^2$, resulting in $\Delta P_{SkL}$ ~ 91 $\mu C/m^2$ excess polarization for the SkL phase. Indeed, in Fig. 5 the magnetic-field dependent polarization curves recorded at temperatures between 10 K and 12 K show two subsequent steps which prove that the cycloidal, SkL, and FM phase exhibit different excess polarizations. This hierarchy of polarization also becomes evident from the isothermal scans in Fig. 5: At $T = 10.5$ K and for increasing magnetic fields, the polarization drops from approximately 87 μC/m$^2$ in the cycloidal phase to 82 μC/m$^2$ in the SkL phase and finally to 73 μC/m$^2$ in the FM state.

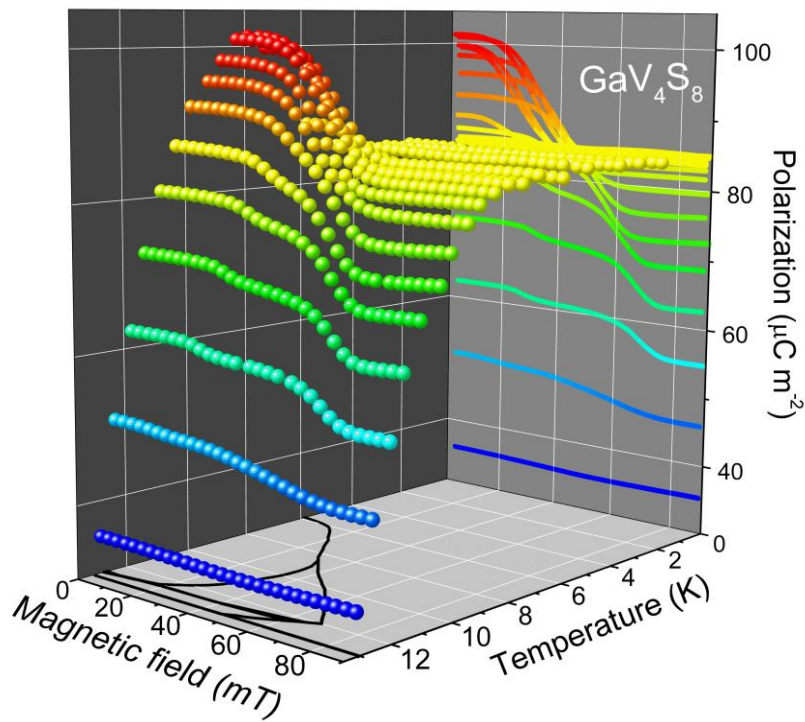

**Fig. 5. Ferroelectric polarization of GaV$_4$S$_8$ determined from magnetocurrent measurements.** Magnetic field dependence of the isothermal polarization measured at various temperatures between 2 and 13 K. Only the excess polarization, Δ$P$, induced by the magnetic ordering is shown. A step-like increase of the polarization at the transition from the cycloidal to SkL and from the SkL to the FM phase can be identified in the $P(H)$ curves measured between 7 and 12.5 K. All the $P(H)$ curves are projected onto the $H$-$P$ plane and all phase boundaries are indicated in the $H$-$T$ plane of the figure. The black lines on the $H$-$T$ plane indicate the same magnetic phase boundaries as in Fig. 2A.

It is interesting to compare the spin-derived polarization in the Néel-type skyrmion host GaV$_4$S$_8$ with that observed in Cu$_2$OSeO$_3$ with Bloch-type skyrmions. In Cu$_2$OSeO$_3$, which is the only insulating material reported to host a SkL state so far, the polarization varies in the range of ±0.5 μC/m$^2$ in the helical and SkL phase (*8,47,48*). This is significantly lower than the polarizations observed in the lacunar spinel compound investigated in the present study. Another important difference is that, while in the helical and FM phase of Cu$_2$OSeO$_3$ the FE polarization continuously increases with increasing external magnetic field and is zero for zero magnetic field (*8*), well developed excess-polarization plateaus are detected in all magnetic phases of GaV$_4$S$_8$, even in the absence of external magnetic fields (see Fig. 5). This fact documents that the excess polarization in GaV$_4$S$_8$ is spin-driven and appears spontaneously at the magnetic phase boundaries, while the polarization in Cu$_2$OSeO$_3$ is induced by external magnetic fields only and follows a quadratic magnetoelectric coupling (*8*).

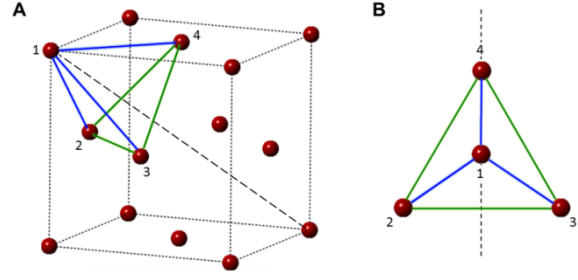

**Fig. 6. Bond symmetry in the rhombohedrally distorted structure of GaV$_4$S$_8$.** (**A**) V$_4$ clusters (represented by red spheres) form an fcc lattice streched along one of the body diagonals (dashed line). Due to this distortion, there are two types of bonds as illustrated for four selected clusters: Intra-plane ones (green lines) between nearest neighbor V$_4$ units within the (111) planes and inter-plane ones (blue lines) connecting V$_4$ units in neighboring (111) planes. (**B**) The four selected clusters viewed from the ⟨111⟩ direction. Each blue bond lies within a mirror plane of the tetrahedron (see dashed line for an example), while green bonds are perpendicular to them.

## Microscopic origin of the spin-driven polarization

Motivated by the distinct values of the spin-driven polarization in the cycloidal, SkL, and FM phase, we analyze its microscopic origin taking into account the rhombohedral symmetry of the lattice and the spin patterns in the three magnetically ordered phases. The building blocks of the magnetic structure, that are the V$_4$ tetrahedra with $S = \frac{1}{2}$, form an fcc lattice elongated parallel to one of the ⟨111⟩ axes. Due to this distortion, the bonds between nearest neighbor V$_4$ units within the (111) planes are not equivalent with those connecting V$_4$ units in neighboring (111) planes as illustrated in Fig. 6. Hereafter, these bonds are referred to as intra- and inter-plane bonds, respectively.

In general, the spin-driven excess polarization, $\Delta P$, can be written as a sum of the polarization of individual bonds being bilinear function of spin components:

$$\Delta P^\alpha = \frac{1}{N} \sum_{\substack{\langle i,j \rangle \\ \beta,\gamma}} \lambda_{i,j}^{\alpha\beta\lambda} S_i^\beta S_j^\gamma , \qquad (1)$$

where the summation goes over all $\beta$ and $\gamma$ spin components for each ⟨$i,j$⟩ bond. Since GaV$_4$S$_8$ is an $S = \frac{1}{2}$ cluster-spin system and the orbital degeneracy is lifted by the rhombohedral distortion, the on-site ($i = j$) polarization terms can be neglected in first order. The form of the magnetoelectric coupling tensors $\lambda_{i,j}^{\alpha\beta\lambda}$ are bond specific and dictated by the symmetry of the bonds (*49,50*). In the experiment only $\Delta P^z$ was detected, where the $z$ axis points along ⟨111⟩. Hence, we will restrict our analysis to the second-rank tensors $\lambda_{i,j}^{(z)\beta\lambda}$ and will omit the $z$ index. (Here we note that due to the presence of FE polarization in the PM phase, $\lambda_{i,j}^{\beta\lambda}$ has the same form, i.e., the same independent non-zero elements, as the exchange coupling matrix $J_{i,j}^{\beta\lambda}$). As shown in Fig. 6, the intra-plane bonds are perpendicular to the mirror planes present in C$_{3v}$ symmetry, while the inter-plane bonds lie within these mirror planes. Correspondingly, the magnetoelectric coupling tensors for a selected intra- and inter-plane bond in Fig. 6b have the forms:

$$\hat{\lambda}_{2,3} = \begin{bmatrix} \lambda^{xx} & \lambda^{xy} & \lambda^{xz} \\ -\lambda^{xy} & \lambda^{yy} & \lambda^{yz} \\ -\lambda^{xz} & \lambda^{yz} & \lambda^{zz} \end{bmatrix}, \quad \hat{\lambda}_{1,4} = \begin{bmatrix} \Lambda^{xx} & 0 & 0 \\ 0 & \Lambda^{yy} & \Lambda^{yz} \\ 0 & \Lambda^{zy} & \Lambda^{zz} \end{bmatrix}. \tag{2}$$

The symmetric and antisymmetric part of the tensors describes the polarization generated by the exchange-striction and spin-current mechanism, respectively (*44,49,51*). The form of $\hat{\lambda}$ matrices for the other bonds can be derived by applying 3-fold rotations.

Next, we calculate the uniform component of $\Delta P_z$ in the magnetically ordered phases using the magnetoelectric coupling tensors described in Eq. (2). Since the wavelength of the magnetic modulation ($2\pi/q$) is more than 20 times larger than the lattice constant $a$ in both the cycloidal and SkL phase (*9*), we use the continuum approximation when describing their spin patterns:

$$\mathbf{S}_{\text{cyc}}(\mathbf{r}) = \frac{1}{2}\left[a_1 \Re\{\mathbf{S}_j e^{i\mathbf{q}_j \mathbf{r}}\} + \text{higher harmonics}\right], \tag{3a}$$

$$\mathbf{S}_{\text{SkL}}(\mathbf{r}) = \frac{1}{2}\left[b_0 \mathbf{S}_0 + b_1 \sum_{j=1}^{3} \Re\{\mathbf{S}_j e^{i\mathbf{q}_j \mathbf{r}}\} + \text{higher harmonics}\right], \tag{3b}$$

$$\mathbf{S}_{\text{FM}}(\mathbf{r}) = \frac{1}{2} c_0 \mathbf{S}_0, \tag{3c}$$

where the propagation vectors of magnetic order are $\mathbf{q}_1 = q(1,0,0)$, $\mathbf{q}_2 = q(-\frac{1}{2}, \frac{\sqrt{3}}{2}, 0)$, $\mathbf{q}_3 = q(-\frac{1}{2}, -\frac{\sqrt{3}}{2}, 0)$ and the corresponding Fourier components are $\mathbf{S}_j = \mathbf{S}_0 - i\mathbf{q}_j/q$ with $\mathbf{S}_0 = (0,0,1)$. In zero magnetic field, we assume that the cycloid has only the fundamental harmonic, i.e. $a_i = 0$ for $i > 0$, which is valid when the magnetic anisotropy is small. The SkL can also be expressed as a Fourier series, where each order contains a superposition of three cycloids whose **q**-vectors sum up to zero. In first order, these are the three fundamental harmonics with $\mathbf{q}_1$, $\mathbf{q}_2$ and $\mathbf{q}_3$. Higher order terms are necessary to keep the spin length constant. The coefficients $a_j$, $b_j$, and $c_0$ depend on the temperature and magnetic field. The uniform component of the polarization can directly be obtained by substituting Eqs. (2) and (3) into Eq. (1) and integrating over the area of the magnetic unit cell of the respective phases:

$$\Delta P_{\text{cyc}} \approx \frac{3}{16} a_1^2 \left[(\lambda^{zz} + \Lambda^{zz}) + \sum_{\alpha}(\lambda^{\alpha\alpha} + \Lambda^{\alpha\alpha})\right], \tag{4a}$$

$$\Delta P_{\text{SkL}} \approx \frac{3}{4} b_0^2 (\lambda^{zz} + \Lambda^{zz}) + \frac{9}{16} \sum_{i=1}^{} b_i^2 \left[(\lambda^{zz} + \Lambda^{zz}) + \sum_{\alpha}(\lambda^{\alpha\alpha} + \Lambda^{\alpha\alpha})\right], \tag{4b}$$

$$\Delta P_{\text{FM}} \approx \frac{3}{4} c_0^2 (\lambda^{zz} + \Lambda^{zz}). \tag{4c}$$

All the polarization terms above come from the exchange-striction mechanism. We assume that contributions from the spin-current mechanism, corresponding to the asymmetric parts of the $\hat{\lambda}$ matrices, can be neglected even in the non-collinear cycloidal and SkL phase due to the slow spatial variation of these magnetic patterns. Additional contributions to the polarization arising from higher harmonics in the cycloidal and SkL state are governed by the strength of the weak exchange anisotropy, $\frac{\delta J}{J} = \frac{2(J^{zz} - J^{xx})}{J^{xx} + J^{zz}} \approx 0.05$, i.e. reduced by the factor of $\delta J/J$ relative to the leading terms in Eqs. (4a) and (4b).

Reflecting the axial symmetry in the rhombohedral phase, the polarization in leading order can be expressed by two parameters of the $\hat{\lambda}$ matrices, namely the sum of their $zz$ elements, $(\lambda^{zz} + \Lambda^{zz})$, and the sum of their traces, $\sum_\alpha (\lambda^{\alpha\alpha} + \Lambda^{\alpha\alpha})$. Next, we determine these parameters based on the measured excess polarizations of cycloidal and FM phase, assuming that the ordered moment in the FM ground state and the low-temperature region of the cycloidal phase is close to S = ½, i.e. $a_1 \approx c_0 \approx 1$. With $\Delta P_{\text{cyc}} = 100$ µC/m² and $\Delta P_{\text{FM}} = 86$ µC/m² we obtain $(\lambda^{zz} + \Lambda^{zz}) = 115$ µC/m² and $\sum_\alpha (\lambda^{\alpha\alpha} + \Lambda^{\alpha\alpha}) = 418$ µC/m². Detailed description of the spin structure in the SkL phase, i.e. the exact values of the $b_n$ coefficients in Eq. (4b), are not available, thus, for the visualization of the spatial dependence of the FE polarization associated with the spin structure of a Néel-type skyrmion we employ the following model of a single skyrmion (*52*):

$$\mathbf{S}_{\text{Sk}}(\mathbf{r}) = \frac{1}{2} \begin{bmatrix} \frac{2\xi x}{x^2 + y^2 + \xi^2} \\ \frac{2\xi y}{x^2 + y^2 + \xi^2} \\ \frac{x^2 + y^2 - \xi^2}{x^2 + y^2 + \xi^2} \end{bmatrix}, \tag{5}$$

where $\xi$ is the effective radius of the skyrmion core. At this radius, the $z$ component of the spins changes sign when moving from the center to the edge of the skyrmion. The spatial dependence of the $z$ component of the spin-driven polarization reads as

$$\Delta P_{\text{Sk}} \approx \frac{3}{2}\left[ \frac{4\xi^2(x^2 + y^2)}{(x^2 + y^2 + \xi^2)^2} \frac{\lambda^{xx} + \Lambda^{xx} + \lambda^{yy} + \Lambda^{yy}}{2} + \frac{(x^2 + y^2 - \xi^2)^2}{(x^2 + y^2 + \xi^2)^2}(\lambda^{xx} + \Lambda^{xx}) \right]. \tag{6}$$

Figure 7 shows the corresponding spatial dependence of the $z$ component of the spin and the polarization. For the magnitude of the magnetoelectric coefficients $\hat{\lambda}$ in Eq. (6), we use the values determined above from the analysis of the cycloidal and FM phases. The characteristic ring like pattern of the $z$ component of the polarization and its strong spatial modulation in the vicinity of the skyrmion core, as high as 25 - 30%, should allow the experimental verification of this nanoscale polarization pattern accompanying the formation of the Néel-type SkL in multiferroic GaV$_4$S$_8$.

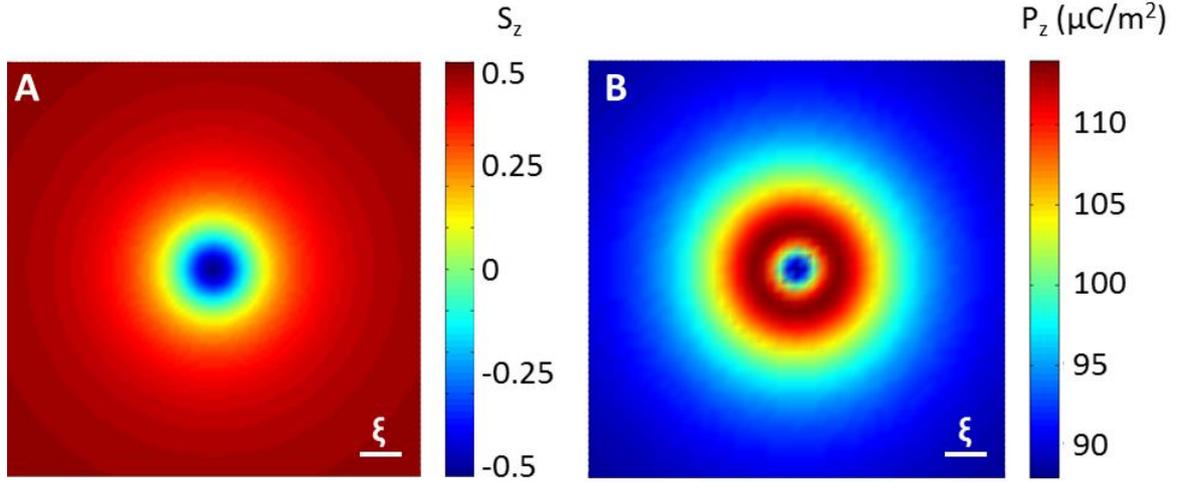

**Fig. 7. Polar dressing of the SkL in GaV$_4$S$_8$.** (**A**) Spatial dependence of the $z$ component of the $S = 1/2$ spins forming the skyrmion and (**B**) spatial dependence of the spin-driven polarization $P_z$, both in the $xy$ plane perpendicular to the vortex core. The radius of the skyrmion core, $\xi$, sets the lateral length scale. The polarization reaches its maximum in a ring-like region around the skyrmion cores (red regions). The outer blue regions in the polarization indicate the FE polarization of the almost collinear spin arrangement at the outer rims of the skyrmion.

## DISCUSSION

In the present work we have identified four different FE phases in the magnetic semiconductor GaV$_4$S$_8$. The large polarization of about 6000 µC/m$^2$ in the FE1 phase, which appears below the Jahn-Teller transition, likely originates from a combination of ionic displacements and redistribution of electron density driven by the ferro-orbital order and is further enhanced when entering the magnetically ordered phases at low temperatures. The excess polarization of the FE2 phase, which is the ground state with collinear FM order, is about 86 µC/m$^2$. The non-collinear spin patterns in the cycloidal and SkL phase further enhance the FE polarization and all multiferroic phases, FE2, FE3, and FE4, can be characterized by three distinct levels of the excess polarization. Although the term multiferroic often refers to the coexistence of ferroelectricity and ferromagnetism only, as realized in the FE2 phase, generally it is used in a broader sense also including more common classes of FE materials with antiferro- or ferrimagnetic order such as the FE3 and FE4 phases of GaV$_4$S$_8$. The four FE phases, out of which three are also magnetically ordered, point to a complex entanglement of charge, orbital, and magnetic degrees of freedom in the lacunar spinel GaV$_4$S$_8$. Most importantly we observe magnetic skyrmions with additional spin-driven ferroelectricity. Model calculations indicate a strong spatial modulation of the FE polarization in the vicinity of the skyrmion core. This donut-shaped polar structure is shown in Fig. 7B.

    We argue that the spin-driven polarization is dominated by the exchange-striction mechanism in each multiferroic phase: i) the spin-dependent orbital hybridization mechanism plays a minor role due to the $S = ½$ spins of the V$_4$ clusters, while ii) the spin-current mechanism is inefficient owing to the slow spatial variation of the spin patterns. The latter means that the Dzyaloshinskii-Moriya interaction generally required for the formation of SkL phases does not significantly contribute to the magnetically induced FE polarization in GaV$_4$S$_8$. We have studied the origin of the magnetoelectric effect on the microscopic level by analyzing

the symmetry of the $V_4$-$V_4$ bonds in the rhombohedral phase and determined the two sets of magnetoelectric coefficients responsible for the spin-driven polarization.

The excess polarization in the SkL phase of $GaV_4S_8$ is almost two orders of magnitude larger than that of $Cu_2OSeO_3$, the only magnetoelectric skyrmion-host material known so far. It is also important to note that in the latter material there is no spontaneous FE polarization, but all polarization is induced by external magnetic fields. The sizable polarization in $GaV_4S_8$ may offer an efficient electric control of the SkL state as already demonstrated in $Cu_2OSeO_3$ (*12,53*), which would be an important step in the development of skyrmion-based memory devices (*3,54,55,56*). This ambitious goal requires, at the first stage, the nano-scale observation of the polar pattern associated with the SkL using local probes such as scanning force and Kelvin-probe microscopy (*57,58,59*).

## METHODS

Polycrystalline $GaV_4S_8$ was prepared by solid state reaction using pure elements of Ga (6N), V (3N) and S (5N). Three subsequent synthesis steps were necessary to obtain full reaction of the starting materials to form the stoichiometric ternary phase. Phase purity after each step was checked by x-ray powder diffraction. The synthesized polycrystals were used as starting material for single crystal growth using chemical transport reactions. The growth was performed in closed quartz ampoules at temperatures between 800 and 850°C utilizing iodine as transport agent. Crushed single crystals have been characterized by x-ray diffraction and were found to be free of any impurity phases. At room temperature we found the correct $GaMo_4S_8$-type structure with F-43m symmetry and a lattice constant $a = 0.966$ nm. Magnetic measurements were performed with a SQUID magnetometer (Quantum Design MPMS XL) in the temperature range from 1.8 K $< T <$ 400 K and in external magnetic fields up to 5 T. The heat capacity was investigated in a Physical Properties Measurement System (Quantum Design PPMS) for temperatures 1.8 K $< T <$ 300 K. The dielectric experiments were performed with a Novocontrol Alpha-Analyzer in a frequency range from 1 Hz to 10 MHz and with a Agilent E4991A impedance analyzer at 1 MHz - 3 GHz. The pyrocurrent has been measured utilizing the Keithley Electrometer 6517A as a function of temperature between 2 and 55 K. In addition we measured the magnetocurrent at temperatures between 2 K and 14 K and in external fields between 0 and 300 mT. In these experiments we used platelet-shaped single crystals of size of 1 mm$^2$ and 0.25 mm thickness. The large (111) surfaces of these samples were contacted by silver paste.

## SUPPLEMENTARY MATERIALS

Supplementary material for this article is available.
Text
Fig. S1. Sample characterization
Fig. S2. Hysteresis behavior of magnetic susceptibility.
Fig. S3. Hysteresis behavior of dielectric constant.
Fig. S4. Dielectric constant up to 300 K.

**Funding:** This research was supported by the DFG via the Transregional Research Collaboration TRR 80: From electronic correlations to functionality (Augsburg/Munich/Stuttgart) and by the Hungarian Research Funds OTKA K 108918, OTKA PD 111756 and Bolyai 00565/14/11.

**Author contributions:** I.K., A.L. and P.L. conceived and supervised the project. V.T. grew the high-quality single crystals. E.R. performed and analyzed the dielectric measurements. S.W. performed and analyzed the specific-heat measurements. S.B. and I.K. performed the calculations. A.L. wrote the paper with contributions from I.K and P.L. All authors discussed the results and commented on the manuscript.

**Data availability:** To obtain data please contact E. Ruff (eugen.ruff@physik.uni-augsburg.de) or P. Lunkenheimer (peter.lunkenheimer@physik.uni-augsburg.de).

**Competing interests:** The authors declare that they have no competing interests.

# Multiferroicity and skyrmions carrying electric polarization in GaV$_4$S$_8$

# Supplementary Materials

E. Ruff,[1]* S. Widmann,[1] P. Lunkenheimer,[1] V. Tsurkan,[1,2] S. Bordács,[3] I. Kézsmárki[1,3], A. Loidl[1]

[1]Experimental Physics V, Center for Electronic Correlations and Magnetism, University of Augsburg, 86135 Augsburg, Germany

[2]Institute of Applied Physics, Academy of Sciences of Moldova, Chisinau MD-2028, Republic of Moldova

[3]Department of Physics, Budapest University of Technology and Economics and MTA-BME Lendület Magneto-optical Spectroscopy Research Group, 1111 Budapest, Hungary

*Corresponding author. E-mail: eugen.ruff@physik.uni-augsburg.de

**Sample characterization**

Figure S1 shows the temperature dependences of the magnetic susceptibility and magnetization measured in the vicinity and below the structural phase transition. In Fig. S1A, the magnetic susceptibility measured at 1 T (open circles, left scale) shows a clear increase at low temperatures indicating the appearance of spontaneous magnetization as expected for a predominantly FM compound. At low temperatures and high external fields, the ordered moment is approximately 0.8 $\mu_B$/f.u., close to the value expected for a spin $S = \frac{1}{2}$ system. The structural phase transition is clearly manifested by a jump of the inverse susceptibility (open squares, right scale) and a sharp peak in the temperature dependence of the heat capacity (Fig. S1B: open squares, left scale), that allow locating the structural phase transition with high precision at $T_{JT}$ = 44.0 K. The magnetic phase transition is also accompanied by a peak in the heat capacity at $T_C$ = 12.7 K.

We estimated the electronic entropy by subtracting the phonon-derived heat capacity. The line in Fig. S1B has been calculated by assuming one Debye term for the acoustic modes and three Einstein terms for the remaining degrees of freedom. A good fit up to room temperature was obtained using a Debye temperature of 145 K and Einstein frequencies corresponding to 240 K, 450 K, and 600 K. By subtracting the phonon part and integrating over the remaining excess heat capacity, we have determined the electronic entropy (open circles in Fig. S1B, right scale). The entropy change upon the magnetic transition is smaller than R ln2 that would correspond to the release of the full spin entropy in the paramagnetic (PM) phase. The entropy just above $T_{JT}$ is approximately 1.43 R ln2. This value is well below R ln2 + R ln3 = 2.58 R ln2, that is the total entropy associated with a single electron per V$_4$ unit occupying a triply degenerate orbital and carrying $S = \frac{1}{2}$. This entropy deficit likely indicates the presence of a dynamic Jahn-Teller effect in the cubic phase, with orbital fluctuations and temporary deformations of V$_4$ units on the local scale (60). The first-order character of the structural phase transition could also hamper a precise determination of the entropy above $T_{JT}$.

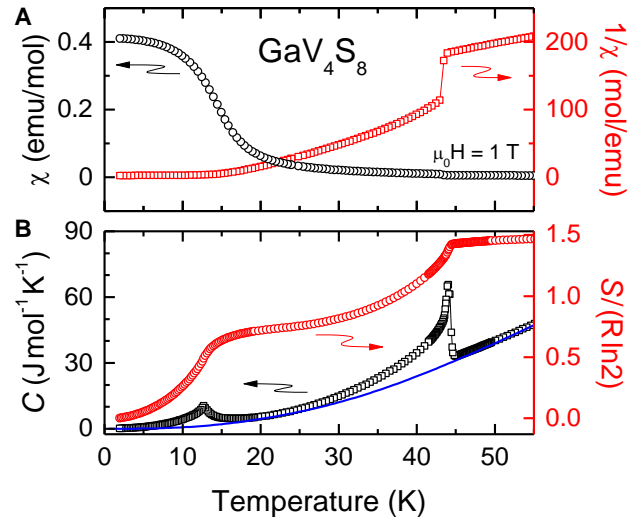

**Fig. S1. Sample characterization.** (A) Magnetic susceptibility $\chi$ = $M/H$ (open circles, left scale) and inverse susceptibility (open squares, right scale) versus temperature measured in external magnetic fields of 1 T along the crystallographic ⟨111⟩ direction. (B) Temperature-dependent heat capacity $C$ (open squares, left scale). The solid line represents an estimate of the phonon contribution to the heat capacity. The electronic entropy $S$, obtained by the integration of the specific heat after subtracting the phonon term, is indicated by open circles (right scale).

**Hysteresis at the Jahn-Teller transition**

Static structural distortions lowering the symmetry from cubic $T_d$ to rhombohedral $C_{3v}$ are expected to occur via a first-order phase transition as odd power terms are allowed in the free energy of a non-centrosymmetric compound (*61*). Indeed, the structural phase transition as seen in the susceptibility and dielectric measurements is accompanied by a hysteresis loop on heating and cooling, similar to the findings in GeV$_4$S$_8$: Figure S2 shows the temperature-dependent magnetic susceptibility as measured under cooling and heating with a rate of 0.5 K/min. At the Jahn-Teller transition, a significant temperature hysteresis is revealed, indicating the first-order nature of the transition.

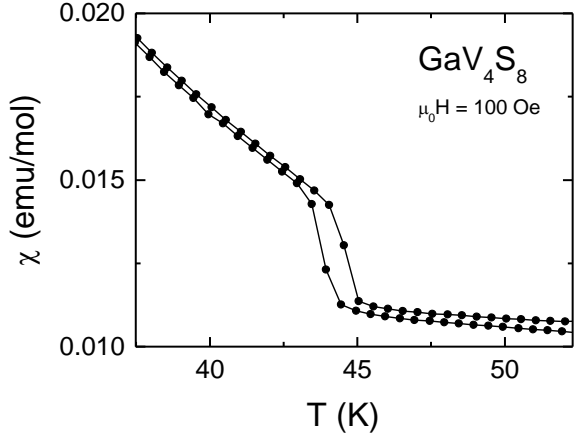

**Fig. S2. Hysteresis behavior of magnetic susceptibility.** Temperature-dependent susceptibility around the Jahn-Teller transition as measured under cooling and heating. A significant hysteresis at the transition is revealed.

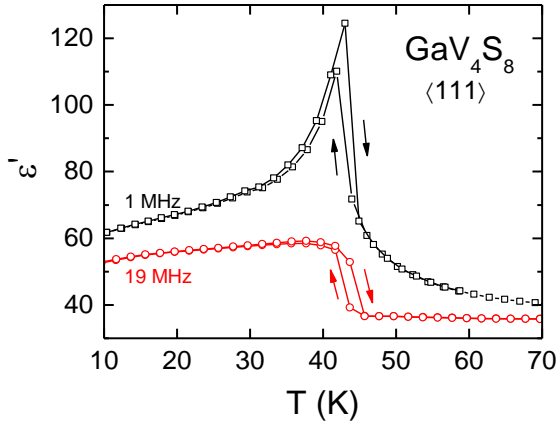

**Fig. S3. Hysteresis behavior of dielectric constant.** Temperature-dependent dielectric constant shown around the Jahn-Teller transition for two frequencies. Results measured under cooling and heating are provided, revealing a significant hysteresis at the transition.

Figure S3 shows the temperature-dependent dielectric constant of GaV$_4$S$_8$ as measured under cooling and heating with a rate of 0.4 K/min. Again, the observed hysteresis at the Jahn-Teller transition indicates its first-order character.

**Maxwell-Wagner relaxation and intrinsic dielectric response**

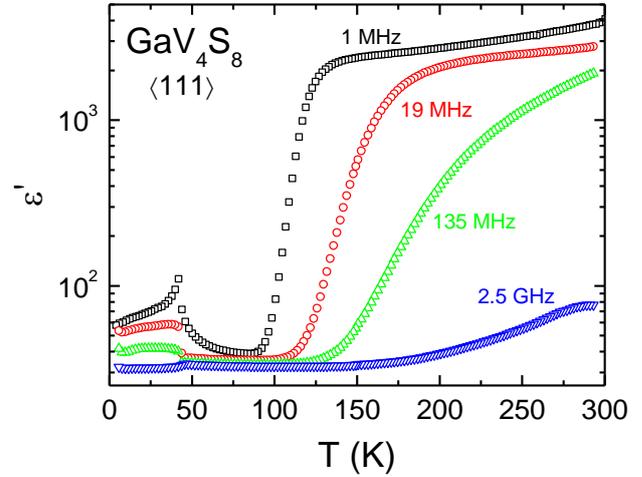

**Fig. S4. Dielectric constant up to 300 K.** Temperature-dependent dielectric constant measured at the same frequencies as shown in Fig. 1C, plotted in an extended range up to room temperature.

Figure S4 shows the temperature dependence of the dielectric constant of GaV$_4$S$_8$ for selected frequencies. Compared to Fig. 1A of the main paper, the data are plotted in an extended temperature range up to 300 K. At high temperatures, huge contributions from an extrinsic Maxwell-Wagner relaxation are revealed, which arise from contact and surface effects and yield values of order $10^3$ at room temperature. As discussed in detail, e.g., in Refs. (*41,42*) of the main text, at low temperatures and sufficiently high frequencies these contributions become suppressed due to the shorting of the capacitance associated with the insulating layers. Fig. S4 demonstrates that in the temperature range around the Jahn-Teller transition ($T_{JT}$ = 44 K) and below, the Maxwell-Wagner relaxation plays no role for the plotted frequencies $\nu \geq$ 1 MHz.

**Determination of the ferroelectric polarization**

The temperature and magnetic field dependence of the ferroelectric polarization have been determined by the integration of pyro- and magnetocurrent data, respectively. We also made attempts to measure polarization hysteresis, $P(E)$, loops in the different FE phases. However, close to the structural phase transition the conductivity contributions are too high. At low temperatures in the magnetically ordered phases we were not able to switch the polarization by external electric fields. Using $P(E)$ techniques, the small excess spin-driven contributions can hardly be detected on the large background from the orbital polarization.